\documentclass{ifacconf}

\usepackage{graphicx}      
\usepackage{natbib}        
\usepackage{subcaption}
\usepackage{multirow}
\usepackage{amsmath}
\usepackage{xcolor}
\usepackage{enumitem}
\setlist[itemize]{noitemsep, topsep=0pt}

\begin{document}
\begin{frontmatter}

\title{European Satellite Benchmark for Control Education and Industrial Training} 

\thanks[footnoteinfo]{This research was funded by the European Space Agency in the frame of the project "Generic Satellite Simulation \& Analysis Tool: Dynamics Modelling, Controls and Validation of Complex Satellite Systems with Guaranteed Robust Pointing Performances": ESA Contract No. 4000141060/23/NL/MGu}

\author[First]{Francesco Sanfedino} 
\author[First]{Paolo Iannelli} 
\author[First]{Daniel Alazard}
\author[Second]{Émilie Pelletier}
\author[Third]{Samir Bennani}
\author[Second]{Bénédicte Girouart}

\address[First]{Fédération ENAC ISAE-SUPAERO ONERA, Université de Toulouse, 10 Avenue Edouard Belin, BP-54032, 31055, France (e-mail: francesco.sanfedino@isae.fr).}
\address[Second]{European Space Agency ESA/ESTEC, Keplerlaan 1, 2201 AZ, Noordwijk, The Netherlands}
\address[Third]{[Retired] European Space Agency ESA/ESTEC}

\begin{abstract}                
To overcome the innovation gap of the Guidance, Navigation and Control (GNC) design
process between research and industrial practice a benchmark of industrial relevance has been developed and is presented. This initiative is driven as well by the necessity to train future GNC engineers and the GNC space community on a set of identified complex problems.
It allows to demonstrate the relevance of state-of- the-art modeling, control and analysis algorithms for future industrial adoption. The modeling philosophy for robust control synthesis, analysis including the control architecture that enables the simulation of the mission, i.e. the acquisition of a high pointing space mission, are provided.
\end{abstract}


\end{frontmatter}

\section{Introduction}
\label{sec:introduction}

As reported by \cite{dennehyverification}, guidance, navigation, and control of aerospace systems is growing more complex together with a need of increased system performance. Over the last decades, academic research developed  several solutions to address this need in modeling, control and validation and verification (V\&V) areas. What is lacking nowadays is: a systematic transfer of these technologies to the industrial world, the scaling up of these advanced
analysis and control algorithms to complex industrial benchmarks and the training of the future control engineers to deal with control problems of industrial complexity. 
High pointing space missions (\cite{Dennehy2018NASAreport}) represent the perfect scenario in which GNC design encounters the hard task of coping with challenging high-level specifications and limitations imposed by the other spacecraft sub-systems (structure, thermal, optics, propulsion, etc.). In order to push the overall system to the limits of achievable performance, it is in fact necessary to predict in a preliminary design phase the worst-case configurations, due both to uncertainty in the system knowledge and mission operation conditions. 
NASA was a pioneer in the field of the integrated modeling philosophy. \cite{young1979interactive} were one of the first to investigate the fully coupled thermal/structure/control analysis problem. This philosophy was then employed in several NASA projects, from the Space Interferometry Mission (\cite{grogan1998}), to the James Webb Space Telescope (\cite{Levine2023}). European Space Agency and industry investigated as well multi-disciplinary model-based system representation for end-to-end pointing performance characterization and optimization. Several missions, from the SILEX experience (\cite{SILEX}) to the recent Euclid Mission (\cite{EUCLID}), benefited of this approach.
The difficulty in practice of correctly conducting worst-case analysis for multi-disciplinary projects lies in their sequential approach, which is commonly followed in industry. Different tools are in fact employed for each area of expertise (GNC, structures, optics, etc.) and worst-case analyses are done in parallel in each field before reaching the final design. This results in many sub-iterations for the exchange of evolving inputs/outputs and trade-offs among different departments and consequent re-validation (\cite{sanfedino2023mono}). A unique user-friendly tool able to incorporate all possible plant configurations (dictated by the particular mission) is then needed to control engineers in order to shortcut this process, synthesize robust control strategies and fast validate system stability and performance.
\begin{figure*}[th!]
    \centering
    \includegraphics[width=.88\linewidth]{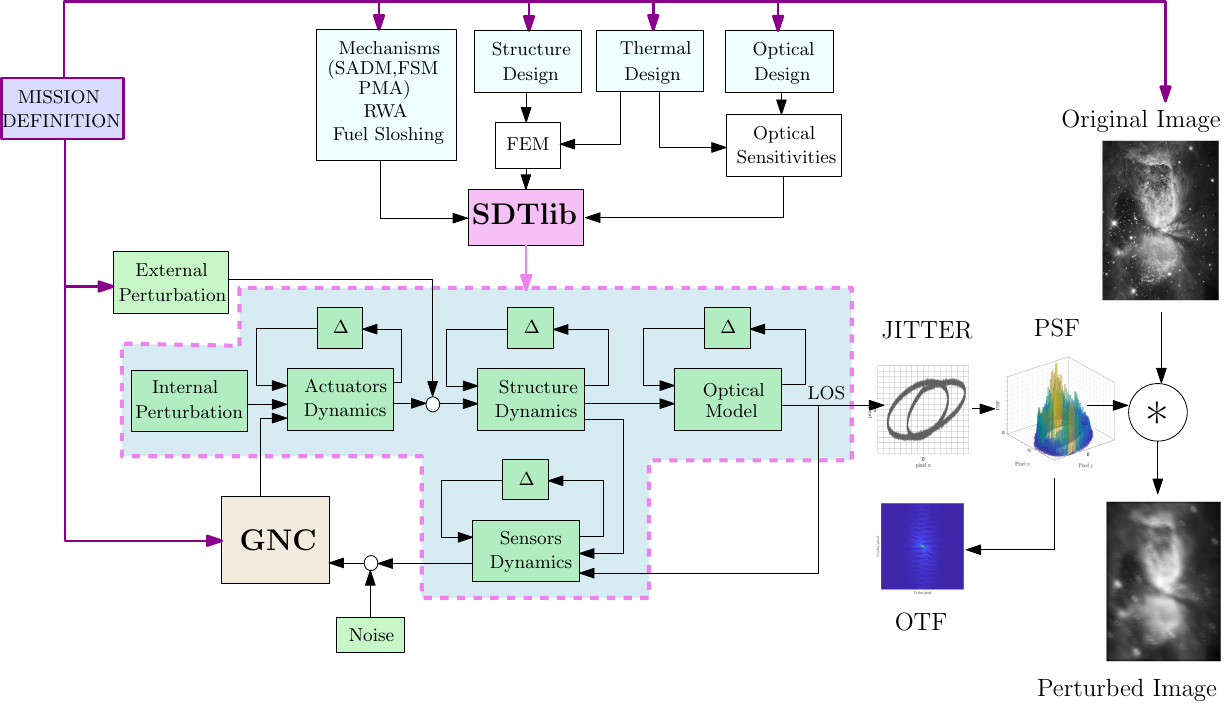}
    \caption{Integrated Design Logic. Hubble Image: \copyright ESA/NASA.}
    \label{fig:IntegratedModel}
\end{figure*}
With this philosophy in mind, the Satellite Dynamics Toolbox library (SDTlib) (\cite{alazard2020satellite,SANFEDINO20239153}) was conceived in order to build a complex multi-body spacecraft in the Two-Input Two-Output Port (TITOP) formalism (\cite{alazard2015two}) as a Linear Fractional Transformation (LFT) (\cite{zhou1995}) model by analytically including any uncertain and variable physical parameter. By having a look to Fig. \ref{fig:IntegratedModel}, structural components coming from analytical models or numerical Finite Element Model (FEM) analysis (imported from NASTRAN) can be directly connected together with mechanism models (like Solar Array Drive Mechanism (SADM), Reaction Wheel Assembly (RWA), Fast Steering Mirror (FSM), etc.), sloshing phenomena, chemical propulsion and analytical optical models. Variation and uncertainties of the overall system can be easily taken into account to model:
\vspace{-5pt}
\begin{itemize}[leftmargin=*]
    \item uncertainties coming from preliminary hypotheses on mission design parameters,
    \item uncertainties coming from not guaranteed FEM validation on Earth in non-operative conditions (no microgravity, simulation of space environment on Earth),
    \item uncertainties coming from misknowledge of sensors and actuators models
    \item varying parameters like orientation of flexible appendages or speed of rotating elements (to correctly take into account gyroscopic effects).
\end{itemize} 
\vspace{-5pt}
For the mechanisms, it is also possible to take into account internal disturbances like static and dynamic imbalances of reaction wheels and harmonic perturbations due to SADM stepper commands, which have to be taken into account for microvibration compensation.
Another feature available in SDTlib is the possibility to build models with the minimum number of occurrences of uncertain and variable parameter by construction and to pre-process the final model (ready for control synthesis and analysis) in order to eliminate non-physical states and reduce the model complexity. This functionalities are particularly interesting when the model has to be used for robust control synthesis (${H}_2 /{H}_\infty$ - synthesis) or formal robust stability/performance analysis ($\mu$ or IQC analysis). 

The benchmark proposed in this paper is constituted of:
\vspace{-5pt}
\begin{itemize}[leftmargin=*]
    \item A scalable analysis and synthesis model (an LFT model of a complex telescope Space mission generated by SDTlib) to enable to perform tasks at various levels of granularity
    \item An equivalent non-linear high-fidelity simulator developed in Simscape and validated with SDTlib for time domain V\&V.
\end{itemize}
\vspace{-5pt}
Note that the user is asked to set the granularity of the model to be considered. The corresponding SDTlib and Simscape models will be then automatically generated. For instance the user can choose to get a model for the synthesis of a coarse pointing performance and choose the sensors/actuators among the ones available and set the degree of fidelity for the non-linear simulator (i.e. for the reaction wheels, one can choose to include static and dynamic imbalances, saturation, friction and/or friction spikes). See Section \ref{sec:mission} for more details.

\begin{figure*}[!th]
  \centering
  \includegraphics[width=.9\linewidth]{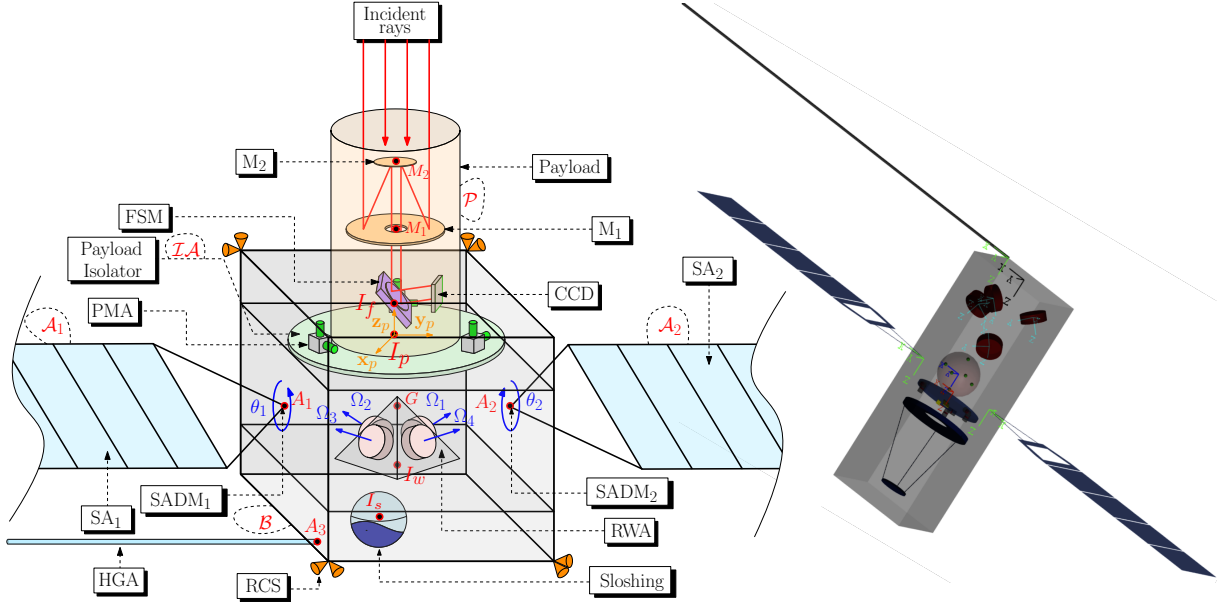}
    \caption{Schematics of the benchmark SPCM (left) and equivalent view in Simscape Mechanical Explorer (right).}
    \label{fig:benchmark_architecture}
\end{figure*}

The optical benchmark can be used for several research problems and applications as listed below (not exhaustive list):
\vspace{-5pt}
\begin{itemize}[leftmargin=*]
    \item End-to-end control synthesis and analysis process for stringent pointing requirements as proposed in \cite{dennehyverification}
    \item Parametric sensitivity analysis as proposed in \cite{ervan2024} in order to reduce the number of most impacting uncertainties on closed-loop stability and performance analysis and better understand the physical meaning of possible worst-case configurations
    \item Deterministic or probabilistic $\mu$-analysis as proposed respectively by \cite{roos2013systems} and \cite{roos2024new}; and with application to flight data in \cite{simplicio2016}
    \item Integral Quadratic Constraints (IQC) analysis as in \cite{veenman2021iqclab} and \cite{PFIFER201536}
    \item Non-linear validation algorithms based on global optimization techniques as proposed by \cite{marcos2013optimization}
    \item Classical fully non-linear Monte-Carlo V\&V campaigns
    \item Multi-disciplinary optimization as proposed by \cite{sanfedino2023robust}
    \item Line-of-sight estimation with different combinations of sensors/actuators as in \cite{SANFEDINO2022107961}
    \item Rigorous analysis (\cite{biannic2010}) of transition between two different control modes (switching)
    \item Exploration of new architectures for actively controlling the spacecraft during Science phase without stopping disturbance sources (like reaction wheels) in order to optimize the science time window as shown in \cite{SANFEDINO2022107961}
    \item Generation of time domain data for any other applications (i.e. data driven control, system identification).
\end{itemize}

\section{Benchmark Architecture}

A schematic of the proposed spacecraft model, the Satellite Pointing \& Control Model (SPCM), is shown in Fig.~\ref{fig:benchmark_architecture}. It includes a central bus $\mathcal{B}$ connected to two rotating flexible solar arrays $\left \{\mathcal{A}_1\,\,\,  \mathcal{A}_2\right \}$, driven both by a SADM, a High-Gain Antenna (HGA) boom respectively at the interface point $A_1$, $A_2$ and $A_3$ and an optical payload system $\mathcal{P}$. The optical payload is composed of a structure connecting the optical elements: the two mirrors $M_1$ and $M_2$, the Charge-Coupled Device (CCD) and a an FSM. This payload is linked to the main body through a passive/active isolator. The isolation assembly ($\mathcal{IA}$) interfaces the payload with the spacecraft bus with the objective of reducing the microvibration transmission from internal disturbance sources, reaction wheels (RWs) and SADMs, to the instruments.
The passive damping typically makes use of visco-elastic materials, springs or hydraulic dampers isolator and it is modelled, in the context of the benchmark, with a 6-DoFs spring-damper system placed at the interface node $I_p$. On the other hand, the payload active isolation can be implemented by relying on a set of inertial proof-mass actuators (PMAs) that directly provide the action to actively counteract microvibrations in the middle-range frequency band as proposed in \cite{SANFEDINO2022107961}. Each reaction wheel is connected to the main bus through a dedicated passive isolator, not shown in the figure for better clarity.
Finally, a tank is connected to the spacecraft hub at the interface node $I_s$ to account for the effect of sloshing on the system's dynamics via an equivalent mechanical model (see \cite{rodrigues2023linear}).

\subsection{AOCS}
The Attitude and Orbit Control System is equipped with:\vspace{-5pt}
\begin{itemize}[leftmargin=*]
    \item A set of 4 RWs wheel in pyramidal configuration
    \item A Reaction Control System (RCS). The RCS is a cluster of thrusters that typically provide higher control authority over the spacecraft attitude with respect to RWs. The use of a Micro-Propulsion System (MPS) can prevent/reduce the impact of microvibrations produced by RWs. The pointing performance that can be achieved depend on the thruster control authority and accuracy which is driven by the minimum impulse bit level; furthermore, these are often operated using Pulse-Width Modulation (PWM) strategies thus introducing non-linearities and limit cycle oscillations. 
\end{itemize}
\vspace{-5pt}
The AOCS sensors suite comprises:
\vspace{-5pt}
\begin{itemize}[leftmargin=*]
    \item A \textit{star tracker (STR)} system for attitude measurements. 
        \item A \textit{gyro (GYR)} system for angular velocity measurements. Two gyro assemblies are made available in the benchmark: a coarse pointing gyro (denominated as GYR-c) and an high performance one (GYR-f) that could be used during fine pointing operations.
        \item A \textit{Fine Guidance Sensor (FGS)} which provides very precise attitude measurements necessary to cope with the tight pointing requirements needed during observations. The FGS is generally positioned near the main instrument and shares its Line-Of-Sight (LOS) in order to limit as much as possible thermo-elastic deformations that can worsen the LOS alignment between the instrument and FGS. The FGS is characterized by a reduced field of view and long image exposure time to provide a sufficient Signal-to-Noise Ratio for star detection.
\end{itemize}

\subsection{Structure Dynamics}
\begin{figure}[!h]
    \centering
    \includegraphics[width=\columnwidth]{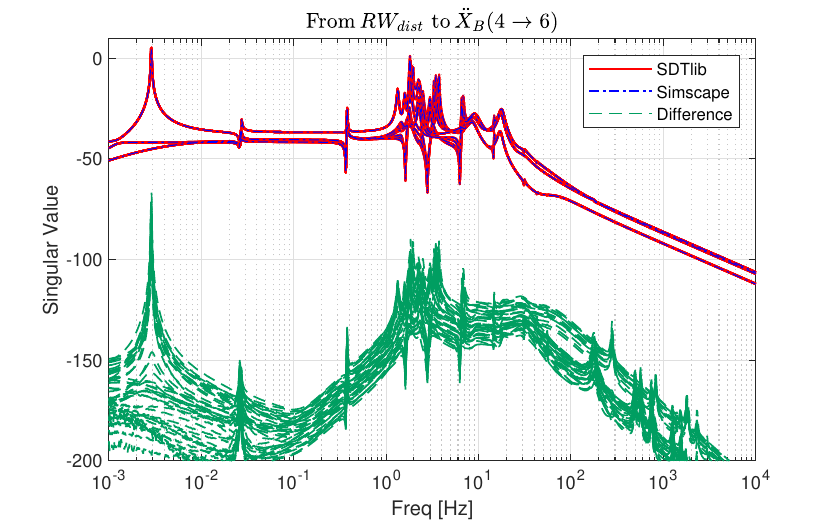}
    \caption{Singular Values of the transfer function from the wheels' harmonic disturbances to the main hub angular accelerations.}
    \label{fig:SAvariation_RWdist2Xdotdot}
\end{figure}

Finite element models of complex structural elements, like the HGA, the solar arrays and the optical payload are analyzed in MSC NASTRAN (see Fig. \ref{fig:telescope}) and directly included in the multi-body SDTlib environment. SDTlib allows also to retrieve the data needed for the Simscape Reduced Order Flexible Solid (ROFS) as shown in \cite{sanfedino2023mono}. All mechanisms are modeled as well in SDTlib and in Simscape environment. Figure \ref{fig:SAvariation_RWdist2Xdotdot} shows for example a comparison of the singular values of the transfer function from the wheel harmonic disturbance to the main hub angular accelerations, where the difference between the two models, depicted by the green lines, is obtained by sampling several configurations of the reaction wheels' speeds.
\vspace{-5pt}

\subsection{Optical Design}

\begin{figure}[!htb]
\begin{minipage}{0.45\columnwidth}
     \centering
     \includegraphics[width=\columnwidth]{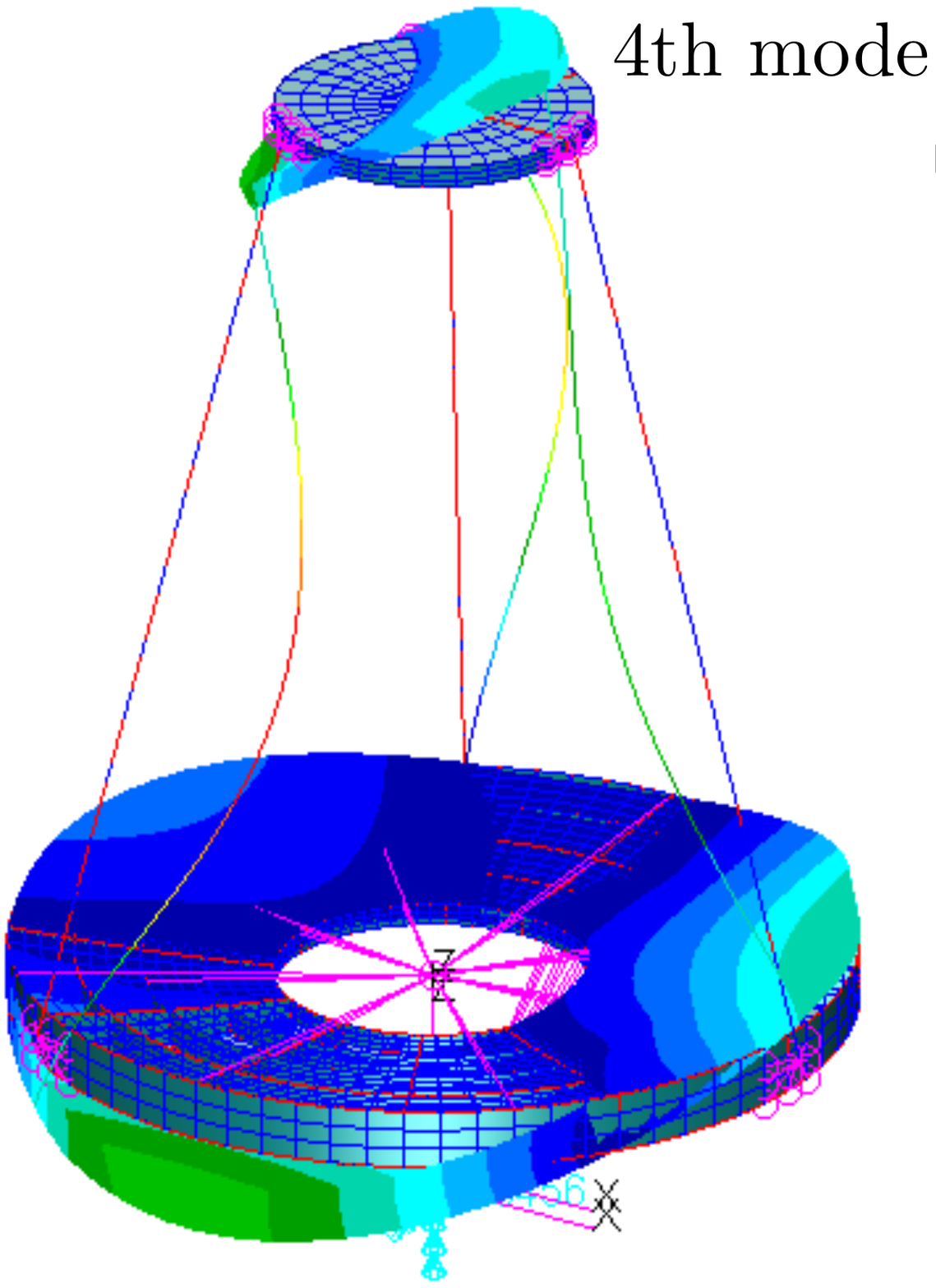}
   \end{minipage}\hfill
   \begin{minipage}{0.45\columnwidth}
     \centering
     \includegraphics[width=\columnwidth]{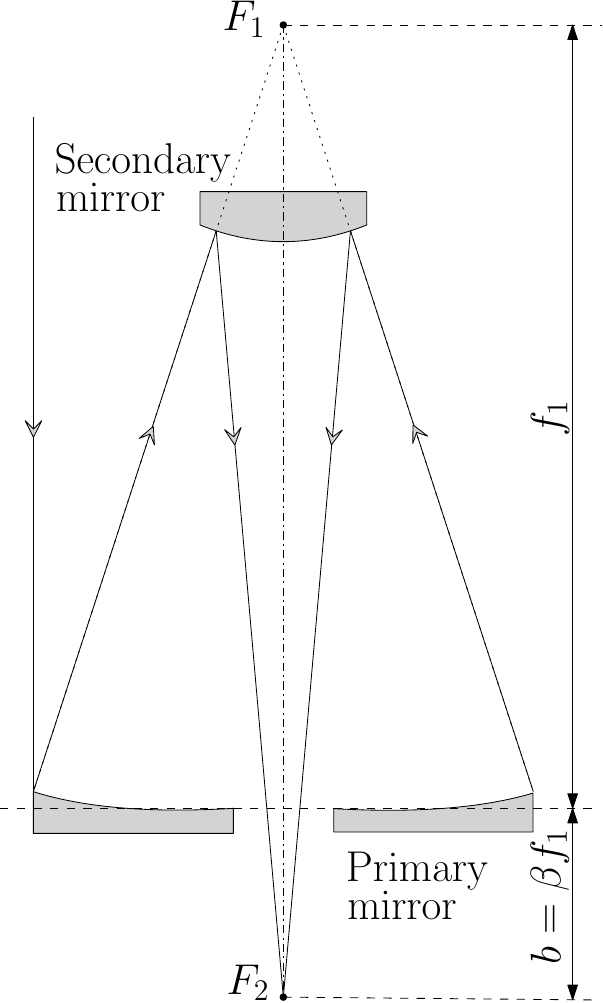}
   \end{minipage}
   \caption{Ritchey-Chrétien two-mirror telescope system: finite element analysis in Nastran (left) and schematics (right)}
   \label{fig:telescope}
\end{figure}

The optical payload chosen for the benchmark is a Ritchey-Chrétien two-mirror telescope system as shown in Fig. \ref{fig:telescope}, where both the primary and secondary mirrors are hyperboloids. The FSM is then placed at half the back focal distance to rotate the
beam towards the detector located on the focal plane.
The ray-tracing algorithm in \cite{redding1991optical} is used to derive the analytical linear optical sensitivities under the paraxial approximation (or small angle approximation). The LOS is then computed by considering an incident ray entering the system along the LOS direction and incident in the centroid of the primary mirror. Since computing the contribution of the optical surfaces’ deformation to the Wave Front Error (WFE) is outside the scope of the benchmark and would require the use of numerical solutions provided by dedicated software, an averaged motion of the primary mirror with a \textit{virtual} point positioned at the vertex of the primary mirror is used to address LOS errors.
\vspace{-5pt}

\subsection{External and internal Disturbances}
\begin{figure}[!h]
    \centering
    \includegraphics[width=.9\columnwidth]{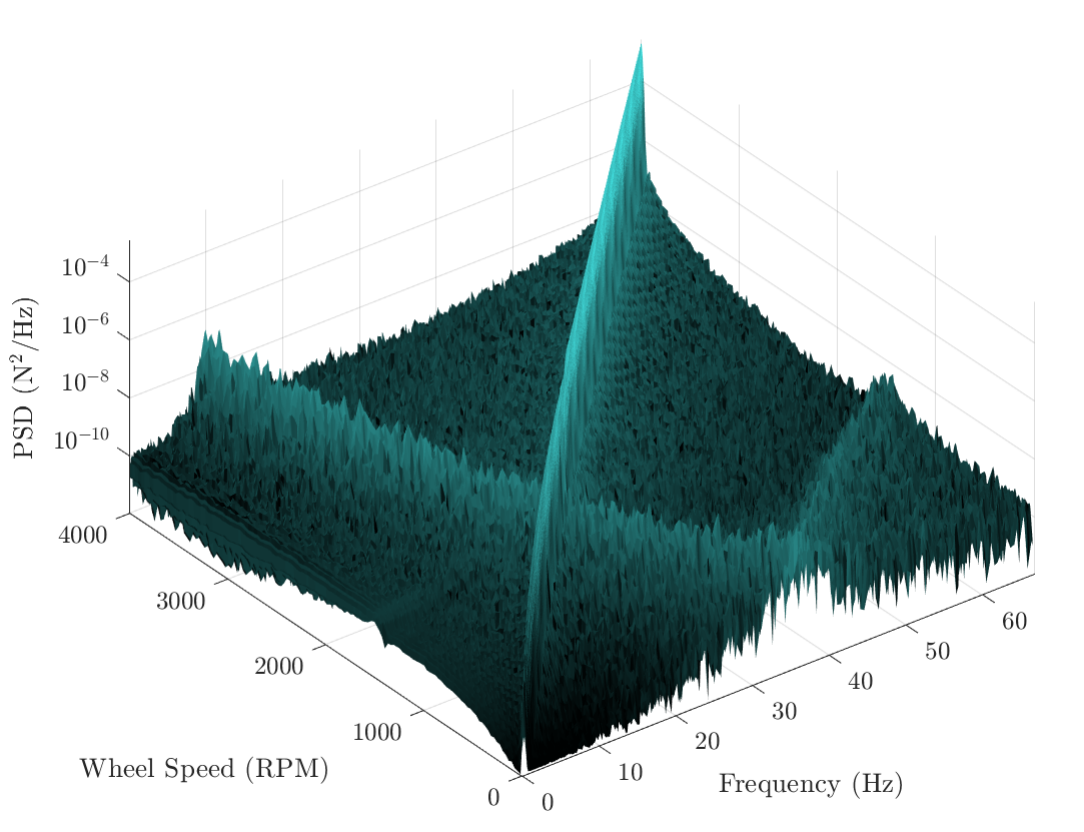}
    \caption{Reaction wheel waterfall plot of the disturbance torque measured in the wheel reference frame. Note the nutation modes of the wheel isolator varing with the wheel speed.}
    \label{fig:watherfall}
\end{figure}

The telescope is supposed to be rotating on a circular orbit, where gravity gradient and solar pressure torques acts on the spacecraft by perturbing its attitude at very low frequency in the AOCS bandwidth.
Internal disturbances coming from the use of SADM and reaction wheels and  are taken into account. In particular as multiple harmonic perturbation due to its microstepping driver is considered at the input of the output shaft holding the solar array as in \cite{SANFEDINO2022108168}. For the reaction wheels, static and dynamic imbalances together with a broadband noise are modeled. In particular the benchmark can be used in order to characterize the influence of each wheel acceleration to the input disturbance in microgravity conditions and taking into account coupling with the rest of the spacecraft structure as shown in Fig. \ref{fig:watherfall}.
In the non-linear simulator Stribek-like friction \cite{olsson1998friction} and friction spikes \cite{ESATRP} of the wheels can be also taken into account.

\vspace{-6.5pt}

\subsection{Propulsion}

The propulsion system is constituted by 12 thrusters in parallelepiped configuration. The  PWM function is build taking into the Minimum Impulse Bit (MIB) and following model presented in \cite{Ieko}. 

\subsection{Stability and Performance metrics}
System stability has to be guaranteed in all phases of the mission. Classical criteria (i.e. SISO gain, phase, delay margins) and more advanced ones (i.e. robust MIMO modulus and disk margins) can be easily imposed.
Pointing performance are imposed on the different phases of the mission (as detailed in Section \ref{sec:mission}) by following ESA standards (\cite{ESAhand}) based on \cite{pittelkau2012}.
Metrics as the Absolute Performance Error (APE), the Relative Performance Error (RPE) and the Performance Drift Error (PDE) are used to define the expected performance on the image quality, dependent on the camera integration time. This pointing performance has to cope as well with stability requirements and performance limitation imposed by sensor/actuator noise and actuation authority. Other criteria can be easily added by the user: control bandwidth, rise time, overshoot, settling time, roll-off, peak gain, discretization, quantization and advanced criteria expressed in various other metrics for MIMO, time varying and nonlinear uncertain systems.

\section{Mission Definition and User Settings}
\label{sec:mission}
\begin{figure}[h!]
	\centering
\includegraphics[width=\columnwidth]{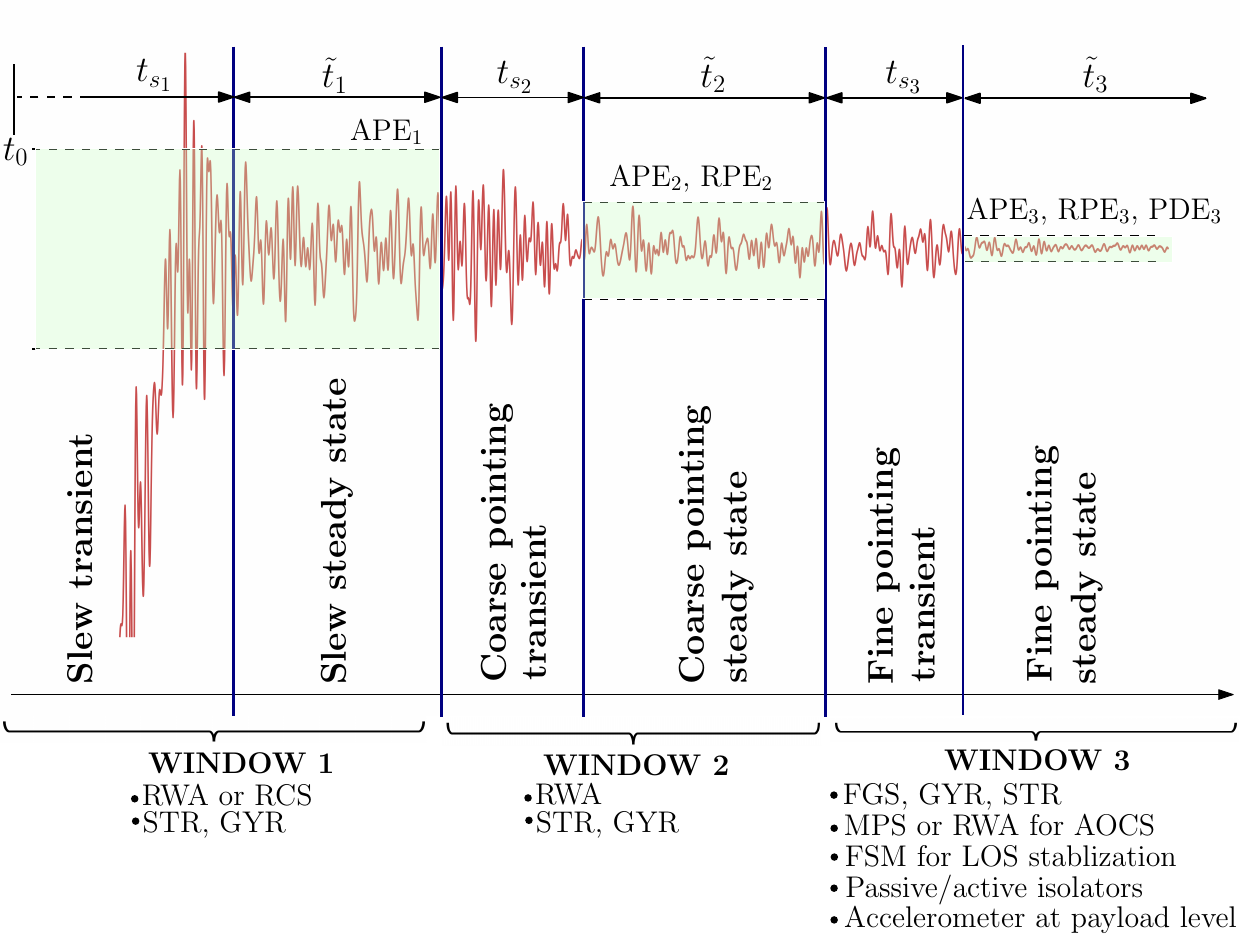}
	\caption{Timeline sequence for the benchmark problem}
	\label{fig:timelineBenchmark}
\end{figure}

Within the various operational modes of the spacecraft, a particular problem is specifically designed as a use case to challenge the future users on the same scenario. This does not limit the use of the present benchmark to other possible investigations as described in Section \ref{sec:introduction}. Figure ~\ref{fig:timelineBenchmark} depicts the timeline for the proposed use case, outlining a sequence of mission phases designed to reach the final science phase which requires precise pointing accuracy and stability to enable observations of a specific target (note that the user is free to select the most suited combination of sensors/actuators proposed in the figure):\\
\textbf{Window 1 - slew transient}: The timeline of the proposed benchmark starts at time $t_0$ and it is initially assumed that the spacecraft performs a slew maneuver to point the payload LOS to the target. Such transient shall be optimized in terms of agility, reduction of the tranquilization time and final attitude error (APE${}_1$). \\
\textbf{Window 1 - slew steady state}: A first APE requirement ($\mathrm{APE}_1$) has to be met during this phase in order to provide the conditions for the mode switching and the transition to the coarse pointing phase. The duration of the slew steady state is considered fixed and equal to $\tilde{t}_1$.\\
\textbf{Window 2 - Coarse pointing transient}: After the specification $\mathrm{APE}_1$ is verified at the end of the slew, the mode switching to the control architecture for coarse pointing control is initiated. This switching involves both the transition from the slew controller to the coarse pointing control scheme and potentially also an AOCS architectural transition (i.e. RCS $\rightarrow $ RWs). This transient ends when the spacecraft fulfill the requirements for coarse pointing steady state operations.\\
\textbf{Window 2 - Coarse pointing steady state}: The coarse pointing phase is envisioned as an intermediate phase in order to provide the sufficient pointing conditions for the initial hand over from star trackers to the FGS and subsequent transition from coarse to fine pointing control architecture. In this phase requirements are provided both in terms of instantaneous and window error indexes ($\mathrm{APE}_2$ and $\mathrm{RPE}_2$). The duration of the coarse pointing steady state is considered fixed and equal to $\tilde{t}_2$.\\
\textbf{Window 3 - Fine pointing transient}: In this phase the hand over to the fine guidance control architecture is initiated. An FGS and high-performance gyro can be activated to provide precise attitude determination, while a second actuator switch at hub level can be envisioned (i.e. from RWs to a MPS). Furthermore, a combination of actuators and sensors at payload level can be incorporated in the control architecture to guarantee the required pointing performance for the optimal execution of the science observation.\\
\textbf{Window 3 - Fine pointing steady state}: The fine pointing phase (science observation phase) presents the most challenging aspects for the AOCS and LOS stabilization system in terms of pointing requirements and complexity of the overall control architecture. Indeed, besides more stringent instantaneous and windowed requirements ($\mathrm{APE}_3$ and $\mathrm{RPE}_3$) with respect to the previous phase, a windowed-stability requirements in the form of Performance Drift Error (PDE) is imposed during science observation ($\mathrm{PDE}_3$). The duration of the fine pointing steady state is considered equal to $\tilde{t}_3$. 
\vspace{-5pt}
The total time for the previous sequence is fixed and a baseline controller is provided to the user, who can directly plug his solution into the simulator. The challenge will be to maximize the science phase, meaning the fine pointing steady state duration $\tilde{t}_3$. This means that the user has to propose a solution that optimizes the slew maneuver and the transitions between modes by coping with all uncertainties of the system and by facing non-linearities like saturations, reaction wheels imbalances, Stribek and spike frictions, SADM microstepping signals, PWM signals for thrusters and all measurement noises. \cite{iannelli2024} extensively outlines all available settings to input data, how to retrieve SDTlib and Simscape models for different control modes, how to set actuators/sensors and how to run a simulation and obtain the automated reporting file.

\section{Conclusion}
A benchmark problem was presented with a threefold goal: 
\vspace{-5pt}
\begin{itemize}[leftmargin=*]
    \item training future GNC engineers with problems of industrial complexity;
    \item trading-off advanced control, analysis and validation tools applied to a realistic satellite control problem by stretching academic tools that often use to collapse on complex large scale problems or do become unduly hard to execute when applied in an industrial context;
    \item transferring to industry the state-of-the-art GNC tools to dramatically reduce the design iterations.
\end{itemize}

\bibliography{ifacconf}             
                                                   







\end{document}